\documentclass[sigconf]{acmart}
\usepackage{xcolor}
\usepackage{wrapfig}

\usepackage{amsmath}
\usepackage{algorithm}
\usepackage{algpseudocode}

\usepackage{multirow}
\usepackage[normalem]{ulem}
\useunder{\uline}{\ul}{}
\usepackage{diagbox}

\usepackage{pifont}

\usepackage{bbding}

\usepackage{hhline}
\usepackage{colortbl}
\usepackage{framed}
\usepackage{mdframed}
\usepackage{amsmath}
\usepackage{array}
\usepackage[caption=false]{subfig}
\usepackage{listings}
\definecolor{darkred}{rgb}{0.6,0.0,0.0}

\definecolor{mycolor1}{RGB}{0, 0, 180}
\definecolor{mycolor2}{RGB}{31, 132, 31}
\newcommand{\sectionspacing}{\vspace{-2.5ex}}

\newcommand{\aftersectionskip}{\vspace{1.5ex}}

\usepackage{microtype}
\usepackage{balance}

\algnewcommand\algorithmicoutput{\textbf{Output:}}
\algnewcommand\OUTPUT{\item[\algorithmicoutput]}

\algnewcommand\algorithmicdefine{\textbf{Define:}}
\algnewcommand\DEFINE{\item[\algorithmicdefine]}

\newif\ifdoubleblind
\doubleblindfalse

\vspace{-10pt}

\def\BibTeX{{\rm B\kern-.05em{\sc i\kern-.025em b}\kern-.08em
    T\kern-.1667em\lower.7ex\hbox{E}\kern-.125emX}}
    
\copyrightyear{2025}
\acmYear{2025}
\setcopyright{cc}
\setcctype{by}
\acmConference[FSE Companion '25]{33rd ACM International Conference on the Foundations of Software Engineering}{June 23--28, 2025}{Trondheim, Norway}
\acmBooktitle{33rd ACM International Conference on the Foundations of Software Engineering (FSE Companion '25), June 23--28, 2025, Trondheim, Norway}\acmDOI{10.1145/3696630.3728500}
\acmISBN{979-8-4007-1276-0/2025/06}

\begin{document}

\title[NeuroStrata:A Neurosymbolic Framework]{NeuroStrata: Harnessing Neurosymbolic Paradigms for Improved Design, Testability, and Verifiability of Autonomous CPS}

%\shorttitle{NeuroStrata}

\ifdoubleblind
  % Double-blind submission: Hide author names
  \author{Anonymous Authors}
\else
  % Camera-ready submission: Show author names
    \author{Xi Zheng}
    \orcid{0000-0002-2572-2355}
    \affiliation{%
      \institution{Macquarie University}
      \city{Sydney}
      \country{Australia}
    }
    \email{james.zheng@mq.edu.au}

    \author{Ziyang Li}
    \orcid{0000-0003-3925-9549}
    \affiliation{%
      \institution{University of Pennsylvania}
      \city{Philadelphia}
      \country{USA}
    }
    \email{liby99@seas.upenn.edu}

    \author{Ivan Ruchkin}
    \orcid{0000-0003-3546-414X}
    \affiliation{%
      \institution{University of Florida}
      \city{Gainesville}
      \country{USA}
    }
    \email{iruchkin@ece.ufl.edu}
    % \IEEEcompsocitemizethanks{
    % \IEEEcompsocthanksitem test
    % }

    \author{Ruzica Piskac}
    \orcid{0000-0002-3267-0776}
    \affiliation{%
      \institution{Yale University}
      \city{New Heaven}
      \country{USA}
    }
    \email{ruzica.piskac@yale.edu}
    
    \author{Miroslav Pajic}
    %\orcid{0000-0001-5282-0658}
    \affiliation{%
      \institution{Duke University}
      \city{Durham}
      \country{USA}
    }
    \email{miroslav.pajic@duke.edu}

    \thanks{
    This work is supported in part by the Australian Research Council (FT240100269, DP210102447, LP190100676), the U.S. National Science Foundation (CCF 2403616, NAIAD 2332744), and the National AI Institute for Edge Computing Leveraging Next Generation Wireless Networks (CNS 2112562).
    %Ivan Ruchkin is supported in part by the NSF Grant CCF-2403616. Any opinions, findings, conclusions, or recommendations expressed in this material are those of the authors and do not necessarily reflect the views of the National Science Foundation (NSF) or the United States Government. 
    }
 %   \author{Armando Solar-Lezama}
    %\orcid{0000-0001-5282-0658}
 %   \affiliation{%
 %     \institution{Massachusetts Institute of Technology}
   %   \city{Cambridge}
   %   \country{USA}
   % }
   % \email{asolar@csail.mit.edu}
    
\fi

\begin{abstract}
Autonomous cyber-physical systems (CPSs) leverage AI for perception, planning, and control but face trust and safety certification challenges due to inherent uncertainties. The neurosymbolic paradigm replaces stochastic layers with interpretable symbolic AI, enabling determinism. While promising, challenges like multisensor fusion, adaptability, and verification remain. This paper introduces \textbf{NeuroStrata}, a neurosymbolic framework to enhance the testing and verification of autonomous CPS. We outline its key components, present early results, and detail future plans.
\end{abstract}
%%
%% The code below is generated by the tool at http://dl.acm.org/ccs.cfm.
%% Please copy and paste the code instead of the example below.
%%


%%
%% Keywords. The author(s) should pick words that accurately describe
%% the work being presented. Separate the keywords with commas.

% \received{20 February 2007}
% \received[revised]{12 March 2009}
% \received[accepted]{5 June 2009}

%%
%% This command processes the author and affiliation and title
%% information and builds the first part of the formatted document.
\maketitle

\vspace{3mm}
\sectionspacing
\section{CONTEXT, MOTIVATION, AND AIMS}

The integration of machine learning (ML) into cyber-physical systems (CPS) has spurred innovations in autonomous vehicles~\cite{Waymo, Tesla, Uber}, delivery drones~\cite{AmazonPrimeAir, GoogleWing, Zipline}, and robotic surgery~\cite{DaVinci, Mazor, Mako}. While ML enhances autonomy, its inherent uncertainty and brittleness—evident in softmax-based classifications and regression-based control—limit the applicability of traditional formal verification, calling for new approaches. \textit{Neurosymbolic methods} address this gap by combining neural learning with symbolic reasoning, enabling co-training via probabilistic logic programming and differentiable reasoning~\cite{cohen2017tensorlog,manhaeve2021neural,li2023scallop,lu_surveying_2024}. Recent advances, including DSL-based program synthesis, support both deterministic and probabilistic behaviors~\cite{parisotto2016neuro,bjorner2023formal,ellis2021dreamcoder}.

Despite progress, key challenges remain: handling unseen deployment data~\cite{fredrikson_learning_2023}, multi-sensor fusion~\cite{strickland_deep_2018}, and the absence of neural component verification and decision-level determinism. These limitations hinder the application of mature formal methods to autonomous CPS. Ensuring safety and liveness requires \textit{systematic co-design} of perception, planning, and control—still lacking in current frameworks.

This paper reviews state-of-the-art neurosymbolic paradigms, using autonomous driving to illustrate limitations in adaptability, sensor integration, component validation, and decision determinism—each critical for reliable CPS operation. To address these, we propose \textbf{NeuroStrata}—a neurosymbolic framework for the design and assurance of autonomous CPSs. It integrates neurosymbolic distillation and LLM-guided test generation for data-driven specification mining, top-down synthesis of hybrid components, and runtime bottom-up adaptation via program induction. This enables dynamic evolution of symbolic programs within perception and control modules, advancing the testability and verifiability of autonomous CPSs.

\begin{figure*}[bt!]
\centering
\includegraphics[width = 0.85\textwidth]
{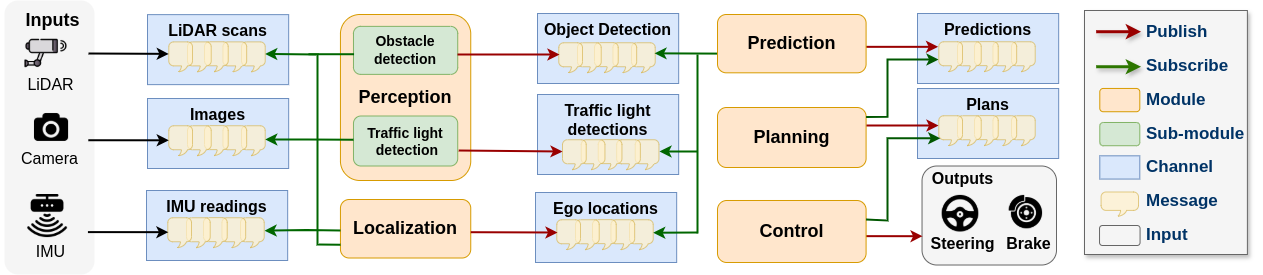}
\vspace{-2mm}
\caption{Motivating system: an industrial-strength software stack for autonomous driving \cite{deng2022scenario}.}
\label{fig:ROSADS}
\end{figure*}

\aftersectionskip

%\vspace{-1.5cm}

\sectionspacing
\section{Motivating System and State-of-the-art}
\label{sec:relatedwork}
State-of-the-art autonomous CPSs, such as autonomous driving systems (ADS), are typically built on middleware frameworks like Robot Operating System (ROS) with proprietary extensions (e.g., Baidu’s CyberRT) to reduce message latency~\cite{zheng2024testing, liang2023rlaga, lou2022testing, deng2021survey, prakash2021multi}. As shown in Figure~\ref{fig:ROSADS}, these systems integrate perception, prediction, planning, and control modules. The perception module processes multi-modal sensor data (e.g., LiDAR, cameras, IMU) for obstacle detection, traffic light recognition, and localization. The prediction module forecasts dynamic object trajectories, while the planning module computes the ego vehicle’s trajectory. The control module generates actuation commands, such as steering and braking. AI components are pervasive across these modules, enabling object detection, trajectory prediction, and real-time control.

\looseness=-1
Recent tools for verifying neural networks, such as NNV star sets~\cite{tran2020verification}, Sherlock~\cite{dutta2019sherlock}, and Reluplex~\cite{katz2017reluplex}, focus primarily on verifying local and input-output properties, which are too narrow to capture system-level safety. %In~\cite{paulsen2022example}, the authors propose learning linear approximations via machine learning to replace hand-crafted bounds, but the guarantees remain pointwise. 
Some efforts build on these tools for system-level safety~\cite{ivanov_verisig_2021,geng_bridging_2024} but are far from verifying real-world systems.
While system-level testing~\cite{deng2022declarative,tian2022mosat,li2020av} and robustness analysis~\cite{tian2018deeptest,deng2020analysis,cai2020real} offer broader insights, they often rely on probabilistic reasoning~\cite{cleaveland_monotonic_2022,badithela_evaluation_2023}, which weakens formal guarantees~\cite{badings_probabilities_2023}.

In parallel, \textit{neurosymbolic paradigms} aim to unify neural and symbolic methods. One way to do so is via differentiable logic programming for producing deterministic yet trainable outputs~\cite{li2023scallop,raghothamandifflog}. Another approach is program induction that synthesizes symbolic programs from limited input-output examples~\cite{parisotto2016neuro,bjorner2023formal,ellis2021dreamcoder}. 

However, neurosymbolic approaches face non-trivial challenges. Differentiable logic programming often requires hard-coded mappings between neural outputs and logical inputs, and its implementation complexity restricts sensor multi-modality and expressive logics needed in autonomous systems. Likewise, program induction relies on fixed domain-specific languages (DSLs) that lack flexibility for sensor fusion or advanced reasoning~\cite{parisotto2016neuro,ellis2021dreamcoder,bjorner2023formal}. Our vision aligns with neurosymbolic reinforcement learning (NS-RL), which combines deep learning with symbolic reasoning for control~\cite{acharya2023neurosymbolic}; however, \textsc{NeuroStrata} proposes a broader architecture-level redesign. It supports formal specification, verification, and adaptation across perception, planning, and control, with NS-RL serving as a possible component at lower levels. \textsc{NeuroStrata} enables end-to-end reasoning, design-time synthesis, runtime validation, and system-level verification over properties expressed in rich logics.

Neurosymbolic distillation is increasingly adopted to unify neural learning with symbolic interpretability. For example, \textit{Automaton Distillation}~\cite{singireddy2023automaton} maps Q-values from deep RL agents into deterministic finite-state automata for policy transfer, while \textit{NUDGE}~\cite{delfosse2023interpretable} distills neural policies into weighted first-order logic rules using neurally guided abstraction and differentiable forward reasoning. In contrast, \textit{NeuroStrata} targets broader system-level specifications across perception, planning, and control, and uniquely integrates such distillation with large language models and \textit{active learning via L*} to extract and refine signal temporal logic specifications, as demonstrated in our prior work~\cite{zheng2024testing}.

\looseness=-1
%In parallel, the neurosymbolic paradigm, which integrates neural networks with symbolic reasoning, offers a promising direction to address these gaps. For instance, Scallop~\cite{li2023scallop} introduces a declarative reasoning framework that integrates neural and symbolic computations, and DiffLog~\cite{raghothamandifflog} provides differentiable logic programming for explainable predictions. While these approaches mark significant progress, they face challenges when applied to autonomous cyber-physical systems (CPS), particularly due to the complexities of integrating multi-modal sensors and addressing dynamic real-world constraints. Likewise, symbolic learning frameworks such as DreamCoder~\cite{ellis2021dreamcoder} and neurosymbolic Program Synthesis~\cite{parisotto2016neuro,ellis2021dreamcoder,bjorner2023formal} show promise but encounter scalability issues, as demonstrated in industrial CPS case studies.
% \tocheckJZ{Ivan: Google is banned here in China and my wall-climbing efforts are not fully effective. Please add the reference for the multimodal neurosymbolic approach for explainable deepfake detection}
%Version: 16 Jan 2025 JZ
Recent work on \textit{multimodal neurosymbolic systems} integrates visual and auditory signals but relies on simplistic symbolic rules, unsuitable for the diverse sensor modalities of autonomous CPS, such as LiDARs, radars, and cameras~\cite{haq_multimodal_2024}. A Neural State Machine (NSM) approach combines vision and language reasoning but suffers from scalability issues, manual scene graph construction, and limited interpretability due to missing source code~\cite{jain2024neuro}. These limitations underscore the need for refined co-design to meet the stringent requirements of autonomous CPS.
%Recent work on \textit{multimodal neurosymbolic systems}, while capable of integrating visual and auditory signals, often relies on simplistic symbolic reasoning rules. These rules are not well-suited to the dynamic and complex requirements of autonomous CPS, which involve diverse sensor modalities, including multiple LiDARs, radars, and cameras~\cite{haq_multimodal_2024}. 
%Another pioneering work integrates vision and language modalities using a Neural State Machine (NSM) for reasoning. However, the underlying scene graph is manually constructed, raising concerns about the generality of the approach. The NSM scalability issues, such as state explosion, and the lack of interpretability due to the absence of concrete source code for reasoning remain significant challenges~\cite{jain2024neuro}. 
%These issues highlight the need for further refinement and co-design to ensure the neurosymbolic paradigm meets the stringent requirements of autonomous CPS.

The above limitations emphasize the pressing need for \textit{deterministic testing and verification approaches} in autonomous CPS, particularly in safety-critical domains like autonomous driving. For instance, perception modules in autonomous vehicles, which rely on uncertain or stochastic processes such as (Bayesian) neural networks for object detection, often fail in ``long tail'' scenarios where the inputs deviate from training data (covariate shift). Predictable and deterministic approaches, incorporating reasoning layers, can adapt to such unseen scenarios by leveraging symbolic logic to ensure robust decision-making and mitigate failures caused by stochastic uncertainties. This adaptability is crucial for guaranteeing safety and reliability across perception, prediction, and planning modules in dynamic and complex real-world environments~\cite{shalev2017formal, seshia2017compositional, zapridou2020runtime}.

\aftersectionskip

\sectionspacing
\section{NeuroStrata: Our Vision for Hierarchical Neurosymbolic Framework for Autonomous Systems}
\label{sec:roadmap}
\begin{figure*}
\centering
% \vspace{-4mm}
\includegraphics[width = 0.75\textwidth]
{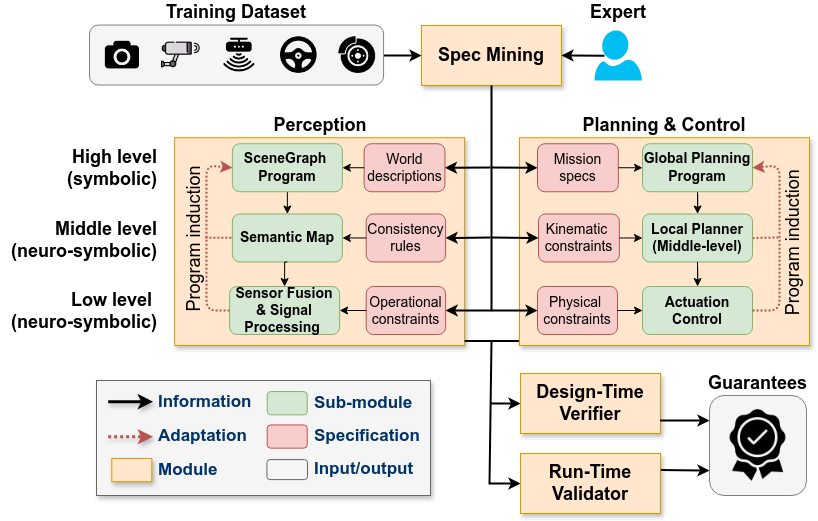}
\caption{Proposed Vision for NeuroStrata: Hierarchical Neurosymbolic Programming for Autonomous CPS}
\label{fig:vision}
% \end{wrapfigure}
\end{figure*}

%Version  5: 11th Jan
%change the first paragraph
\looseness=-1
To address the challenges of designing, testing, and verifying autonomous CPS, we propose a new \textbf{neurosymbolic framework}, \textbf{NeuroStrata}, tailored to the unique requirements of such systems. As shown in Figure \ref{fig:vision}, NeuroStrata combines neural adaptability with symbolic reasoning to enforce formal specifications across hierarchical DSLs that capture underlying safety and liveness properties. The framework structures \textit{Perception} and \textit{Planning \& Control} capabilities into high-level (symbolic-only) and middle- and low-level (neurosymbolic) modules. It ensures runtime reliability and adaptation via a two-phase process: \textit{top-down synthesis}, propagating symbolic specifications to neurosymbolic modules, and \textit{bottom-up adaptation}, where neurosymbolic outputs refine symbolic programs.

% \begin{wrapfigure}[16]{r}{0.75\textwidth}

%Version 4: 9th Jan
%To address the challenges of testing and verifying autonomous CPS, we propose a new \textbf{neurosymbolic framework}, \textbf{NeuroStrata}, tailored to the unique requirements of such systems. As shown in Figure \ref{fig:vision}, NeuroStrata integrates scalable frameworks like Scallop~\cite{li2023scallop} with hierarchical DSLs to ensure formal specification enforcement and adaptability at all levels. This framework applies to both \textit{Perception} (context-awareness) and \textit{Planning \& Control}, employing a two-phase process: \textit{top-down synthesis (design-time)} and \textit{bottom-up adaptation (runtime)}.

\textbf{Modules.} 
%At design time, \textit{Spec Mining}, which is built on top of existing neurosymbolic distillation \cite{
%singireddy2023automaton,blazek2024automated,abir2024neuro},to extract formal safety and liveness specifications from training datasets. 
At design time, \textit{Specification Mining}, built on neurosymbolic distillation \cite{singireddy2023automaton,blazek2024automated,delfosse2023interpretable}, extracts formal safety and liveness specifications from training datasets.
To cover more diverse safety and liveness violations and out-of-distribution scenarios beyond existing training data, we leverage recent work using large language models to analyze multi-modal sensor data \cite{zheng2024testing,deng2023target}, such as front-facing cameras in vehicles, to generate additional real-world crashes and unusual cases from various angles.
These specifications are propagated hierarchically across the system. In the perception stack, a high-level \textit{Scene Graph} encodes semantic relationships and interactions between objects (e.g., ``pedestrian crossing road''), represented as differentiable, adaptable programs that can be verified using formal tools like theorem provers. The middle-level \textit{Semantic Map} encodes spatial and semantic information such as road layouts and drivable areas, ensuring consistency with the scene graph via symbolic rules. The low-level \textit{Sensor Fusion and Signal Processing} integrates multi-modal sensor data (e.g., LiDAR, cameras, GPS) while enforcing constraints on accuracy and consistency, leveraging neurosymbolic reasoning for fusion and processing. Similarly, the planning and control stack follows a hierarchical structure. The high-level \textit{Global/Mission Planner} synthesizes deterministic programs to achieve overall system objectives, verified with formal methods such as theorem proving. The middle-level \textit{Local Planner} generates short-term trajectories that align with global plans while adapting to local changes, guided by symbolic reasoning. The low-level \textit{Actuation Control} converts trajectories into control commands (e.g., steering angle, throttle) and ensures compliance with constraints using runtime verification techniques.

\textbf{Specifications.} 
For \textit{perception}, high-level specifications govern system-wide context awareness, such as ensuring that pedestrians and vehicles do not spatially overlap in the scene graph or that all objects adhere to semantic relationships. Middle-level specifications enforce localized consistency, such as aligning lane boundaries with the semantic map and ensuring that detected objects are positioned correctly within the road layout. Low-level specifications address operational constraints, such as maintaining sensor fusion accuracy within a 0.1-meter error margin and ensuring consistent integration of multi-modal sensor data. For \textit{planning/control}, high-level specifications ensure system-wide safety and mission compliance, such as requiring the vehicle to remain within designated route bounds throughout its journey. Middle-level specifications enforce trajectory-level constraints, such as avoiding obstacles within a 2-meter radius or maintaining smooth transitions between trajectory points. Low-level specifications govern detailed actuation control, such as keeping the steering angle within physical limits and ensuring the stability of throttle and braking in response to control inputs. 
These hierarchical specifications for perception and planning/control ensure an integrated and reliable system design.
%Together, these hierarchical specifications for perception and planning/control enable a robust, consistent, and safe system design.

\textbf{Adaptation.} 
At runtime, symbolic reasoning and neural components interact through a bi-directional feedback loop. Neural modules (e.g., sensor fusion and perception) generate scene-level representations, which are continuously validated against symbolic specifications such as semantic constraints and temporal logic. 
These neural outputs trigger updates to symbolic programs (e.g., the scene graph program and global planning program) via differentiable program induction (see Figure~\ref{fig:vision}). 
This progressively evolves symbolic components in response to new observations. 
In turn, symbolic planners guide neural policy adaptation and ensure specification compliance at all layers. 
This closed neuro-symbolic loop supports runtime behavior that is both adaptive and formally grounded.
%During runtime, NeuroStrata dynamically adapts its perception and planning modules to real-time data while maintaining formal specification compliance. For perception, sensor data flows upward through the hierarchy, where outputs from the low-level sensor fusion are validated against middle-level semantic map constraints, and updates propagate to the high-level scene graph. It evolves dynamically using differentiable program induction, compacting, and adapting specifications as needed. For planning and control, high-level mission planners adjust strategies based on changing conditions, while differentiable and adaptable control programs refine global plans and compact themselves in response to system data. Middle- and low-level components, such as local planners and actuation control, remain guided by symbolic reasoning to ensure safety and alignment with global objectives. This integration enables simultaneously adaptable and formally validated behavior throughout the system.

\textbf{Guarantees.} 
NeuroStrata ensures reliability through a hybrid validation framework. High-level deterministic programs, such as scene graphs and mission planners, are validated using formal verification tools like model checking and theorem proving. Middle- and low-level neurosymbolic components, such as semantic maps and sensor fusion, are guided by symbolic constraints and validated using white-box testing, runtime monitors, and error propagation analysis (e.g., approximate reachability verification~\cite{geng_bridging_2024} and conformance checking~\cite{habeeb_approximate_2024}). Together, this framework bridges the gap between deterministic high-level programs and adaptive, data-driven neuro-components, thus providing formal guarantees across all three levels of the hierarchy.

\aftersectionskip

\sectionspacing
\section{EARLY RESULTS AND FUTURE PLAN}
\label{sec:earlyresults}
% \tocheckJZ{ZiYang, the SE community would greatly appreciate seeing some earlier results demonstrating how the neurosymbolic paradigm can improve the testability and verifiability of autonomous CPS. Now that the design is clearer, could you provide a quick summary of your LASER work? Specifically, where you used LLMs to extract a customized spatial-temporal specification language from captions, and applied this to weakly supervise the generation of spatial-temporal scene graphs? This could effectively address the question of how neurosymbolic-extracted DSLs can support training scene graphs that align with formal specifications. RQ1 is designed for this purpose, while I will explore answers for RQ2.}
We conducted a preliminary case study to explore the key \textbf{Research Question (RQ)}: ``Can neurosymbolic reasoning complement neural network training to better align with specifications?'' We also outline future directions, challenges, and proposed solutions.
%We conducted a preliminary case study to investigate a key \textbf{Research Question (RQ)}: ``can neurosymbolic reasoning complement neural-network training to align with underlying specifications?''. We also outline our future plans, along with the potential challenges and proposed solutions. %detail our future plans along with potential challenges and solutions. %Similarly, an initial exploration was performed for \textbf{Research Question 2 (RQ2)}: ``can program induction enable effective neurosymbolic reasoning?''
%Our preliminary findings are promising, showing how a neurosymbolic approach not only enhances interpretability but also ensures alignment with formal specifications, particularly for tasks such as scene graph generation and spatial-temporal reasoning. These results align well with our roadmap and provide a strong foundation for future exploration.

\subsection{Specification-Aligned Training}
\label{sec:symbolicReasoning}
\begin{figure}
    \centering
    \includegraphics[width=\linewidth]{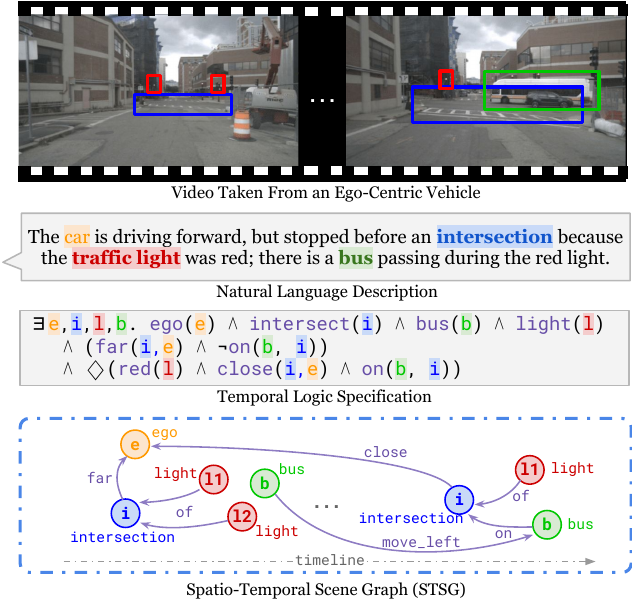}
    \caption{Aligning STSG with natural language description via temporal logic specifications.}
    \label{fig:laser-illus}
\end{figure}

Our recent study, LASER~\cite{huang2024laser}, explores aligning neural perception models with formal specifications via differentiable neurosymbolic reasoning.
The task is to infer spatio-temporal scene graphs (STSGs) from egocentric videos, with outputs constrained by spatio-temporal logic specifications.
Unlike pure symbolic methods, which lack perceptual grounding, our approach enables backpropagation through a differentiable symbolic engine—implemented via Scallop—thus training neural components not only from data but also from logical constraints.
Figure~\ref{fig:laser-illus} illustrates a traffic scenario where a natural language description is formalized into a temporal logic formula involving operators such as $\exists$, $\wedge$, $\neg$, and $\Diamond$.
This enables the neural model to be trained through symbolic specification alignment rather than direct supervision, alleviating the need for the hard-to-obtain fine-grained labels.
We evaluate on OpenPVSG~\cite{yang2023panoptic}, Something-Something~\cite{goyal2017something}, and MUGEN~\cite{hayes2022mugen}, showing that our method improves downstream performance and interpretability, validating the promise of top-down, specification-guided synthesis.
For instance, LASER reaches $53.98\%$ Recall@10 on object relation prediction, superceding fully-supervised baseline (VPS-Conv~\cite{yang2023panoptic}) by $+30.55\%$.
When fine-tuned on a large-scale weakly-labeled dataset with 87K video-caption pairs in the LLaVA-Video dataset~\cite{zhang2024llavavideo}, out-of-domain LASER model is capable of reaching on-par performance ($-0.52\%$) with in-domain performance, further exemplifying its generalizability.

\subsection{Advancing \textsc{NeuroStrata}: A Six-Step Plan}

To operationalize \textsc{NeuroStrata}, we propose a six-step roadmap that addresses key challenges in specification-driven learning and adaptation for AI-based CPS.

I. \textit{Diverse training dataset generation} will leverage model-based scenario synthesis using large language models, as demonstrated in~\cite{zheng2024testing, elmaaroufi2024scenicnl}, and validated on real systems in~\cite{deng2023target}.
This approach mitigates covariate shift by generating multi-modal, edge-case–rich training and testing datasets, grounded in physical realism and enhanced with domain-specific constraints.

II. \textit{Domain-specific languages (DSLs)} will be co-designed with robotics and autonomy experts to express hierarchical specifications across sensing, planning, and control.
These DSLs serve as a bridge between formal methods and system engineering, ensuring usability while enabling precise constraint encoding for downstream reasoning and synthesis.

III. \textit{Specification mining via neurosymbolic distillation} will extract logical properties from training data and LLM queries. To manage hallucinated or noisy outputs, we will integrate symbolic-neural hybrid techniques and active learning to iteratively refine mined specifications for correctness and interpretability.

IV. During \textit{design-time synthesis and verification}, symbolic and neurosymbolic modules will be synthesized from DSL-defined specifications and verified using compositional and abstraction-based methods.
Parallel verification and modular analysis will allow scalability to high-dimensional state spaces, ensuring robust deployment in complex CPS environments.

V. For \textit{runtime adaptation and validation}, \textsc{NeuroStrata} will employ differentiable program induction~\cite{ellis2021dreamcoder} to update symbolic programs on-the-fly. To ensure computational feasibility, we will build on the efficient runtime monitoring infrastructure in~\cite{zheng2016efficient}, which uses low-overhead logging and an aspect-oriented programming paradigm to implement scalable runtime verifiers. This setup supports dynamic refinement of neural-symbolic components with minimal latency, addressing both real-time constraints and covariate shift at runtime.

VI. For \textit{industrial deployment}, \textsc{NeuroStrata} will transition from simulation-based testing in CARLA, and AirSim-based environments to real-world integration with autonomous vehicles, uncrewed aerial vehicles, and robotics platforms. We have initiated research agreements with industry partners in these domains to support this transition. Our upcoming Shonan Seminar (No.~235) and Dagstuhl Seminar (No.~202501048) will further accelerate industry engagement and foster collaborative development and deployment of neurosymbolic CPS solutions.

\aftersectionskip
%\section{A case study}
%\input{Casestudy}
%\label{sec:casestudy}

\sectionspacing
\section{CONCLUSION}
\label{sec:conclusion}
This paper investigates the potential of neurosymbolic paradigms in the design, testing, and verification of autonomous CPS. We introduce \textbf{NeuroStrata}, a framework that supports top-down synthesis of symbolic and neurosymbolic components for perception and control, and bottom-up adaptation of symbolic programs for real-time decision-making. Preliminary results demonstrate neural alignment with specifications. We discuss key challenges and implementation strategies, aiming to bridge theory and practice in autonomous CPS verification.
\aftersectionskip

\clearpage

\balance
\bibliographystyle{ACM-Reference-Format}
\bibliography{main}
%\printbibliography

\end{document}